\documentclass[%
 reprint,
 amsmath,amssymb,
 aps,
]{revtex4-2}

\usepackage[utf8]{inputenc}

\usepackage{subfigure}
\usepackage{capt-of}
\usepackage{here}
\usepackage{soul}
\usepackage{graphicx}
\usepackage{dcolumn}
\usepackage{bm}
\usepackage{xcolor}

\begin{document}

\preprint{APS/123-QED}

\title[]{Dynamic Multiple Nanoparticle Trapping Using Metamaterial Plasmonic Tweezers}

\author{Domna G. Kotsifaki}
\author{Viet Giang Truong}
\email{Corresponding author: v.g.truong@oist.jp}
\author{S\'{i}le Nic Chormaic}

\affiliation{Light-Matter Interactions for Quantum Technologies Unit, Okinawa Institute of Science and Technology Graduate University, Onna, Okinawa, 904-0495, Japan}

\date{\today}

\title{Dynamic Multiple Nanoparticle Trapping using Metamaterial Plasmonic Tweezers}

\date{\today}

\begin{abstract}
Optical manipulation has attracted remarkable interest owing to its versatile and non-invasive nature. However, conventional optical trapping remains inefficient for the nanoscopic world. The emergence of plasmonics in recent years has brought a revolutionary change in overcoming limitations due to diffraction and the requirements for high trapping laser powers. Among the near-field optical trapping cavity-based systems, Fano resonant optical tweezers have a robust trapping capability. In this work, we experimentally demonstrate sequential trapping of 20~nm particles through the use of metamaterial plasmonic optical tweezers. We investigate the multiple trapping via trap stiffness measurements for various trapping positions at low and high incident laser intensities. Our configuration could be used as a light-driven nanoscale sorting device under low laser excitation. Our results provide an alternative approach to trap multiple nanoparticles at distinct hotspots, enabling new ways to control mass transport on the nanoscale.
\end{abstract}


\maketitle


Light interactions with metallic nanostructures demonstrate unique properties which have drawn the attention of researchers, giving rise to an emergent field called plasmonics. Currently, the study of artificial metallic nanostructures~\cite{R7} has greatly expanded the range of electromagnetic properties exhibited by naturally occurring materials. Because metamaterials are produced by structuring their unit elements on a spatial scale that is smaller than the light wavelength, they occupy a unique niche between photonic crystals and regular materials. Numerous applications of metamaterials include superlenses~\cite{R2}, spintronic~\cite{R4}, cloaking~\cite{R5} and lasing spasers~\cite{R6}.  A state-of-the-art design, which presents a resonant magnetic response is the asymmetric split-ring resonator (ASR)~\cite{R9,R10,R37}. In this structure, both the radiant and subradiant modes are supported due to the structural symmetric breaking and the interference between these two modes generates a Fano resonance peak~\cite{R7}. Fano resonant metamaterials enable strong selective enhancement of the electromagnetic fields in their immediate vicinity, thereby considerably boosting the interaction of light with matter placed in close proximity to the metamolecule. Hence, even with a weak perturbation in the electromagnetic environment of the metamaterial, the scattering properties can be significantly altered~\cite{R45}. Owing to their physical properties, Fano resonant nanosystems can enhance the performance for nanophotonic applications such as Raman spectroscopy for molecular detection~\cite{R12,R13} and optical trapping~\cite{R37}. 

Optical tweezers~\cite{R18,R19} are an attractive technique for direct manipulation of particles in a free-solution environment. Benefiting from the rapid development of localized surface plasmon  technology, plasmonic optical trapping can break the optical diffraction limit and further enhances the near electromagnetic field leading to strong optical gradient forces~\cite{R14}. As a result, plasmonic optical tweezers (POTs)~\cite{R35,R76}, have led to numerous fundamental studies and technical applications. For example, double nano-aperture-assisted POTs have been applied to proteins and other biomolecules~\cite{R17}, providing a new way to better understand their interactions in real time. While much previous work has studied single-particle trapping, the trapping of multiple particles has attracted great interest with  applications in physics and biology~\cite{R22,R70}.
Furthermore, by arraying the metallic features in nanostructures~\cite{R22,R29,R32,R37,R69,R75}, many trapping sites can be activated at the same time, permitting simultaneous analysis of several nanoparticles, locally trapped in well-defined positions of the plasmonic devices. 
 The ability to trap multiple particles in periodic arrays suggests the possibility of creating synthetic nanomaterials~\cite{R71,R72,R29} and may enable on-chip biological trapping and analysis. For example,  sequential optical trapping of a single 30~nm particle has been demonstrated using a double nano-aperture array with 44~nm connecting nanoslots~\cite{R22}. The authors achieved  sequential trapping of single nanoparticles using an on-resonance laser with an intensity of 0.51~mW/$\mu$m$^{2}$ ~\cite{R22}. Recently, we demonstrated 20~nm polystyrene (PS) nanoparticle trapping on an array of asymmetrical split ring (ASR) nano-apertures~\cite{R37}. A relatively large normalized trap stiffness of 8.65~fN/nm/mW at the near-resonant condition was achieved~\cite{R37}. This high trap stiffness enabled trapping at low incident intensities and leads to a variety of potential applications in chemistry and biosensing.

In this paper, we use a Fano resonant, metamaterial optical tweezers to tightly trap 20~nm polystyrene (PS) particles in succession. We perform trap stiffness measurements at several plasmonic hotspots and trapping wavelengths. We show that the nanoparticle is transferred towards the plasmonic nanostructure and trapped at a plasmonic hotspot,  which is not necessarily that which provides the strongest optical forces. By switching off and on the trapping laser beam, the trapped particle moves to the adjacent trapping position due to its Brownian motion. This implies that our plasmonic configuration may operate as a plasmonic sorting device. We show that the experimental trap stiffness and the theoretical optical force curves versus trapping wavelength, for each trapping position, follow a similar trend, indicating the dominance of the Fano resonant mechanism on the trapping performance at laser intensities up to~1.10~mW/$\mu$m$^{2}$. By multiple particle trapping without forming any clusters, we foresee that our plasmonic device will open up new avenues for nanoparticle manipulation in a lab-on-chip environment. 

\begin{figure}[ht]
\centering
\includegraphics[width = 85 mm, height = 55 mm]{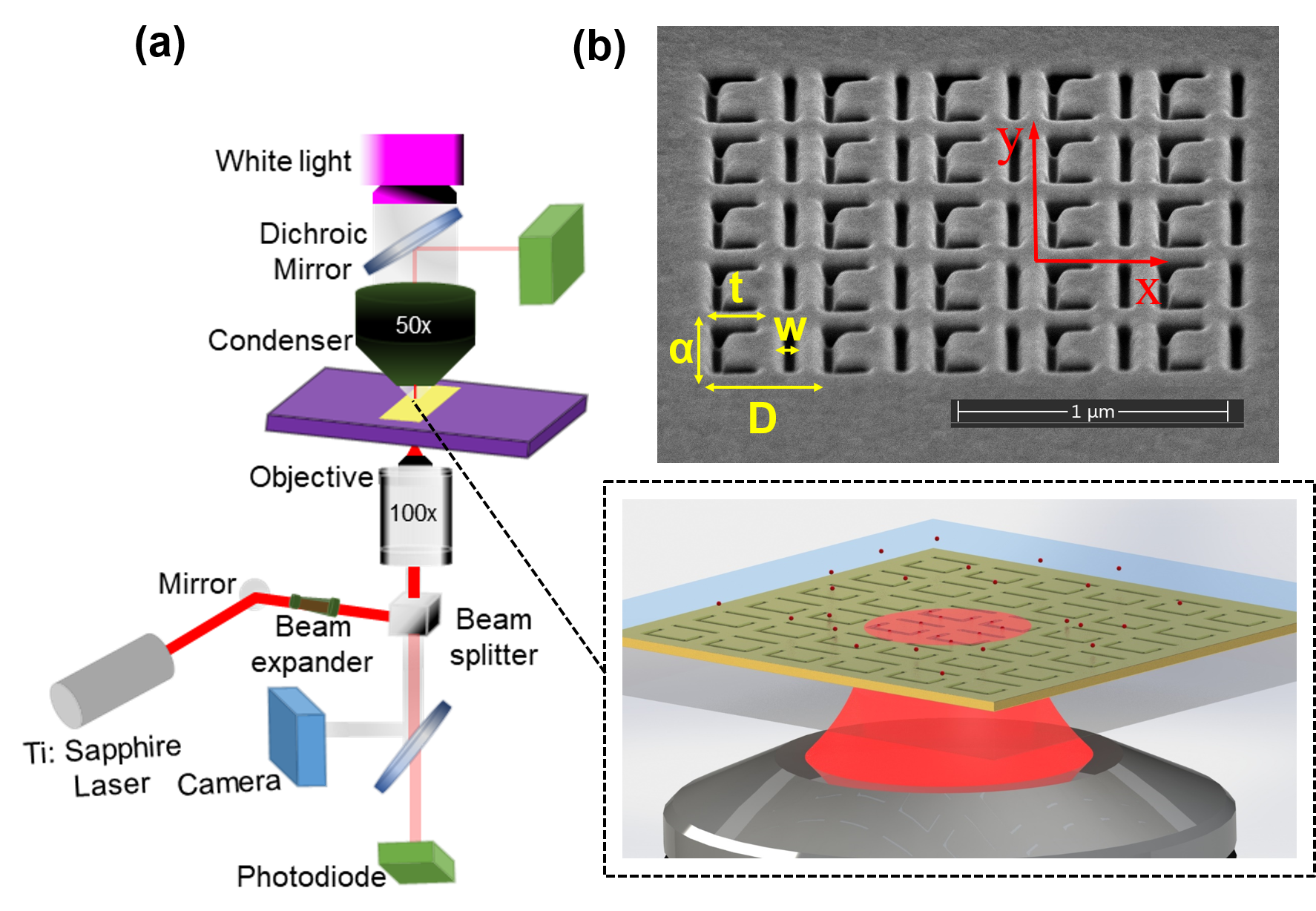}
\caption{\label{Fig.1} (a) Schematic illustration of the experimental setup, showing a zoomed in image of the trap substrate region. (b) Scanning electron microscope (SEM) image, viewed at $52^{\circ}$ from the surface normal of the metamaterial. The geometrical dimensions of each metamolecule unit are: \textit{D} = 405\,~$\pm\,~2$~nm, vertical slit \textit{$\alpha$} = 310\,~$\pm\,~4$~nm,  horizontal slit \textit{t} = 164\,~$\pm\,~3$~nm, and slit width \textit{w} = 44\,~$\pm\,~2$~nm~\cite{R37}.}
\end{figure}

For the optical trapping measurements, we used a custom-built trapping set-up, with a tunable, continuous-wave Ti:Sapphire laser as the trapping laser (Figure~\ref{Fig.1}(a)). The laser beam was focused by a high numerical aperture oil-immersion microscope objective lens (NA = 1.25, 100$\times$) near the metamaterial. Dielectric PS particles of 20~nm diameter were diluted into heavy water with a 0.0625\% volume concentration. A small amount of detergent (Tween-20) with 0.1\% was added to the particle solution and the final solution was sonicated to further prevent the formation of aggregates. To detect a trapping event, the transmitted laser light was collected through a 50$\times$ objective lens and sent to an avalanche photodiode (APD).  We monitored the APD signal via a 100~kHz sampling rate data acquisition card. We fabricated the metamaterial structure using focused ion beam (FIB) milling through a 50-nm gold thin film. Figure~\ref{Fig.1}(b) shows a scanning electron microscope (SEM) image of an array of 5$\times$5~metamolecules with a periodicity along both $x$ and $y$ of 405~nm.The metamaterial used for the trapping experiments consisted of an array of 17 (\textit{x}-direction) and 15 (\textit{y}-directions) ASR metamolecules. From the transmission and reflection spectra of the metamaterial, which were measured using a  microspectrophotometer (MCRAIC 20/30PV), we calculated the absorption spectrum to determine the Fano resonance peak at 928~nm~\cite{R37}. 

To gain insight into the physical mechanism underlying the trapping of nanoparticles at specific trapping positions, we performed finite element method simulations using the COMSOL Multiphysics software package. The geometric parameters used in the simulations were those obtained from SEM images (Figure~\ref{Fig.1}(b)). We calculated the near-field (\textit{E}-field) distribution, the  optical forces, and the potentials when the metamolecule unit was polarized along the \textit{x}-direction (Figure~\ref{Fig.1}(b)). 
The optical forces on the trapped nanoparticle were calculated based on the time-averaged Maxwell's Stress Tensor method~\cite{R38,R68}. We assumed that a 20~nm PS nanoparticle was positioned at the strongest near-field intensity region in the water side of the open nano-aperture area where the equilibrium position exists. Potentials were calculated by integrating the optical forces along the \textit{z}-axis.

\begin{figure}[htp]
\centering
\includegraphics[trim={0.3cm 0cm 0cm 0.1cm},clip,width=0.51\textwidth]{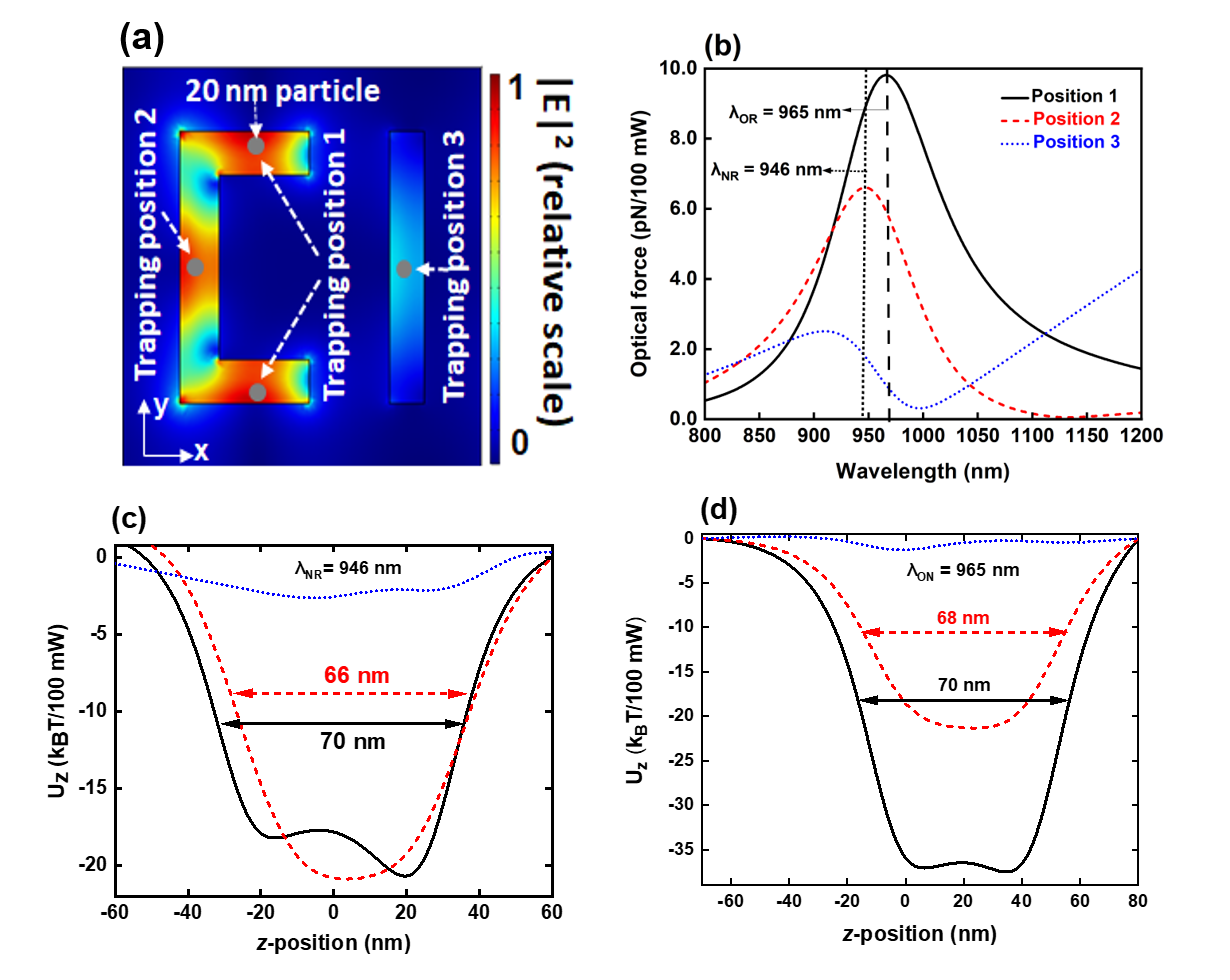}
\caption{\label{Fig.2} (a) Near-field (\textit{E}-field) distribution of an ASR unit at a simulated resonance of 965~nm for the \textit{x}-direction.  The three possible trapping positions are indicated. (b) The optical forces exerted on a 20~nm PS particle at the three trapping positions. Potential well for each trapping position along the \textit{z}-axis for (c) off-resonance illumination at 946 nm and (d) on-resonance conditions at 965 nm. In our model, the plane at \textit{z}~=~0 is the boundary between the gold (Au) film and the SiO$_{2}$ interface. The nanoaperture pattern was cut into the glass substrate to a depth of 20~nm to provide a short distance over which the nanoparticle can move close to the gold/glass interface.}
\end{figure}

Figure~\ref{Fig.2}(a) shows the simulated enhancement of the electric field distribution on the \textit{xy}-plane for a resonant wavelength of 965~nm. When the ASR unit is illuminated by light, the strong light-matter coupling creates three plasmonic hotspots, as shown by the red regions of Figure~\ref{Fig.2}(a). The theoretical Fano resonance peak occurs at 965 nm.  It is associated with excitation of circulating currents oscillating in the two types of nanoapertures, that is the \textit{C}-type and \textit{l}-type. Such asymmetric modes yield magnetic dipole moments oriented normal to the array plane, oscillating synchronously in all metamolecules~\cite{R73, R74}. Figure~\ref{Fig.2}(b) shows the theoretical optical force exerted on a 20~nm PS particle for the three possible trapping positions as a function of the incident trapping wavelength. Figure~\ref{Fig.2}(c) and (d) show the trapping potentials for off- and on-resonance conditions, respectively.  From Figure~\ref{Fig.2}(b), the nanoparticle appears to prefer to be trapped at the position where the near-field exerts a maximum force upon it (trapping Position 1, Figure~\ref{Fig.2}(a)). Moreover, the geometry of our device means that the nanoparticle can also be trapped at positions where the optical force is less strong compared to Position 1 (Figure~\ref{Fig.2}(a)-(c)). Furthermore, we observe a red-shift of the optical force  at trapping Position 1 compared to trapping Positions 2 and 3. This red-shift  originates from the Fano resonance contributions of each individual hotspot~\cite{R37}. For \textit{x}-polarized excitation, the antisymmetric mode, which associated with the absorption resonance, can dominate compare to the usual symmetric one. As a consequence, the strength of the induced currents can reach high values at trapping Position 1 compared to Positions 2 and 3, thereby leading to the red-shift. We determine the full-width-at-half-maximum (\textit{FWHM}) of the trapping wells in the \textit{z}-direction for each trapping position when on-resonance (Figure~\ref{Fig.2}(d)) and we calculate the respective theoretical trap stiffnesses to be $\kappa_{1}$~=~4.16~fN/nm, $\kappa_{2}$~=~1.16~fN/nm, and $\kappa_{3}$~=~0.28~fN/nm.      

\begin{figure}[ht]
\includegraphics[width = 88 mm, height = 92 mm]{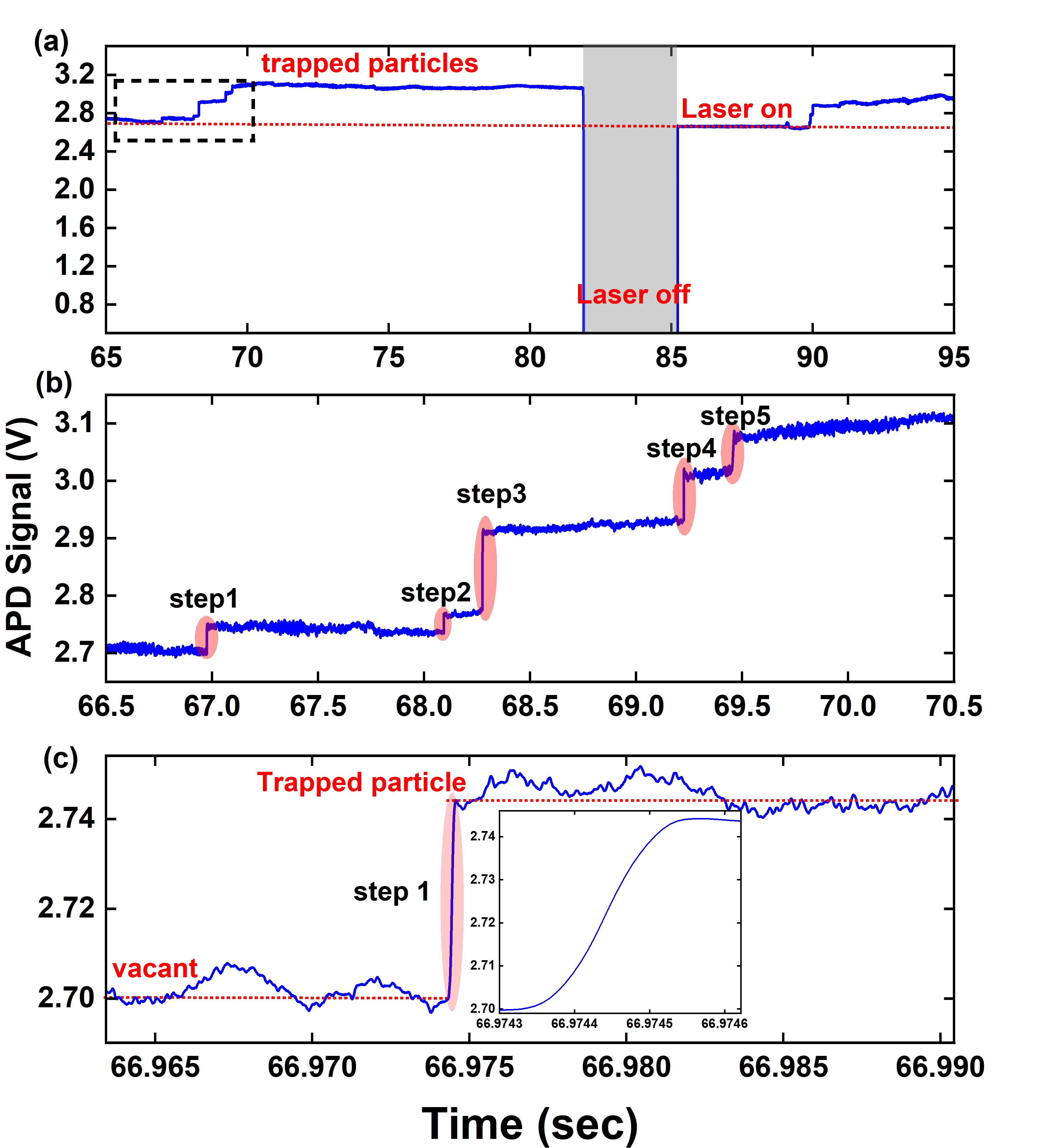}
\setlength\abovecaptionskip{-5pt}
\caption{\label{Fig.3} (a) Laser transmission through the ASR array, showing plasmonic trapping events for 20~nm PS particles recorded over 30~sec. The incident laser intensity is 0.53~mW/$\mu$m$^{2}$ at the sample plane and the trapping wavelength is 923~nm. (b) Expanded view of several trapping events of black dashed rectangular box in (a). (c) The expanded time trace  of the first trapping event in (b), representing a time interval of 0.3~ms.}
\end{figure}

Figure~\ref{Fig.3}(a) shows a trace of the transmission signal recorded on the APD as a function of time through an array of ASR metamaterials in a solution of 20~nm PS beads. Zoomed-in images of Figure~\ref{Fig.3}(a) reveal multiple step-like jumps, typically within 4~sec, indicating optical trapping events (see Figure~\ref{Fig.3}(b)). At 65~sec the trapping laser is turned on and the first PS nanoparticle is trapped after two seconds (67~sec). After one more second (68~sec), several nanoparticles are sequentially trapped at plasmonic hotspots of the metamaterial device. We observe five distinct steps, relative to the vacant level, of the sequential nanoparticle trapping. The cycle of the particle trapping and releasing process is repeated by blocking and unblocking the trapping laser beam. After re-illumination, we notice similar step-like signals, indicating clearly the repeatability of the multiple nano-trapping process. Due to the geometrical dimensions of each ASR metamolecule, we confirm that the trapped nanoparticle remains trapped at the same hotspot for the period of time during which the trapping laser is on, preventing the transportation of the nanoparticle to a neighboring hotspot.

To investigate  multiple trapping using the Fano resonant metamaterial plasmonic tweezers, we measured the trap stiffness for each trapping event, at several trapping positions, for a low (0.55~mW/$\mu$m$^{2}$), and a high (1.10~mW/$\mu$m$^{2}$) incident laser intensity. 
The overdamped Langevin equation~\cite{R24} can be used to describe the particle motion within the trap:
\begin{equation}
    \frac{dx(t)}{dt}=\frac{\kappa_{tot}}{\gamma}x(t)+(\frac{2k_{B}T}{\gamma})^{1/2}\zeta(t)
\end{equation}
where \textit{$\gamma$} is the viscous damping,~\textit{x(t)} the displacement of the particle from the equilibrium point,~\textit{$\kappa_{tot}$} the total trap stiffness,~\textit{$\zeta$(t)} is the white noise,~\textit{$k_{B}$} is the Boltzmann constant and~\textit{T} is the temperature. Using an exponential fit of the trapping region, the trap stiffness is determined from~\cite{R24}:

\begin{figure*}[ht]
\centering
\begin{minipage}[c]{0.44\linewidth}
\includegraphics[trim={1.7cm 0.5cm 2.8cm 1.8cm},clip, width=1\textwidth]{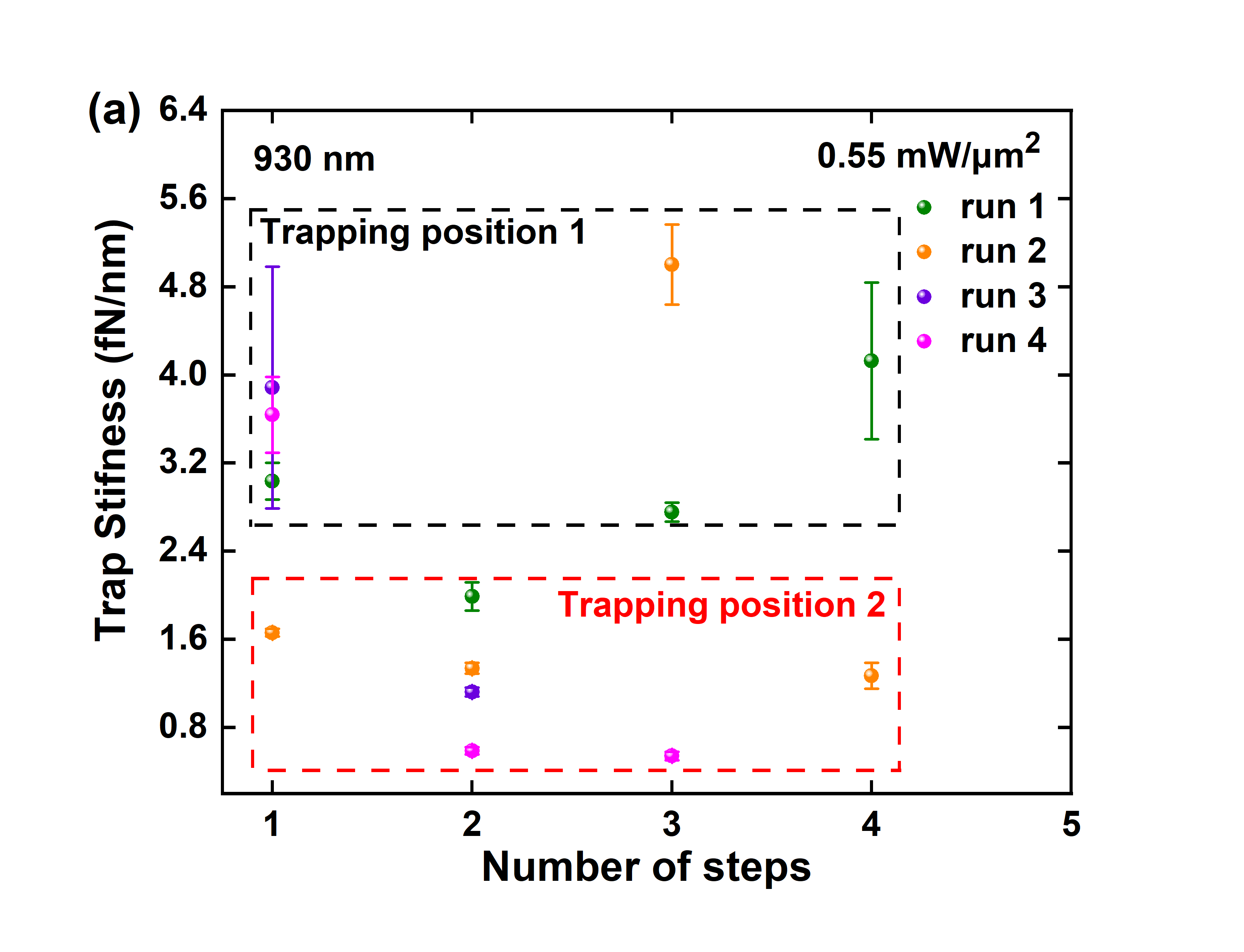}
\end{minipage}
\begin{minipage}[c]{0.44\linewidth}
\includegraphics[trim={1.7cm 0.2cm 2.8cm 2cm},clip, width=1\textwidth]{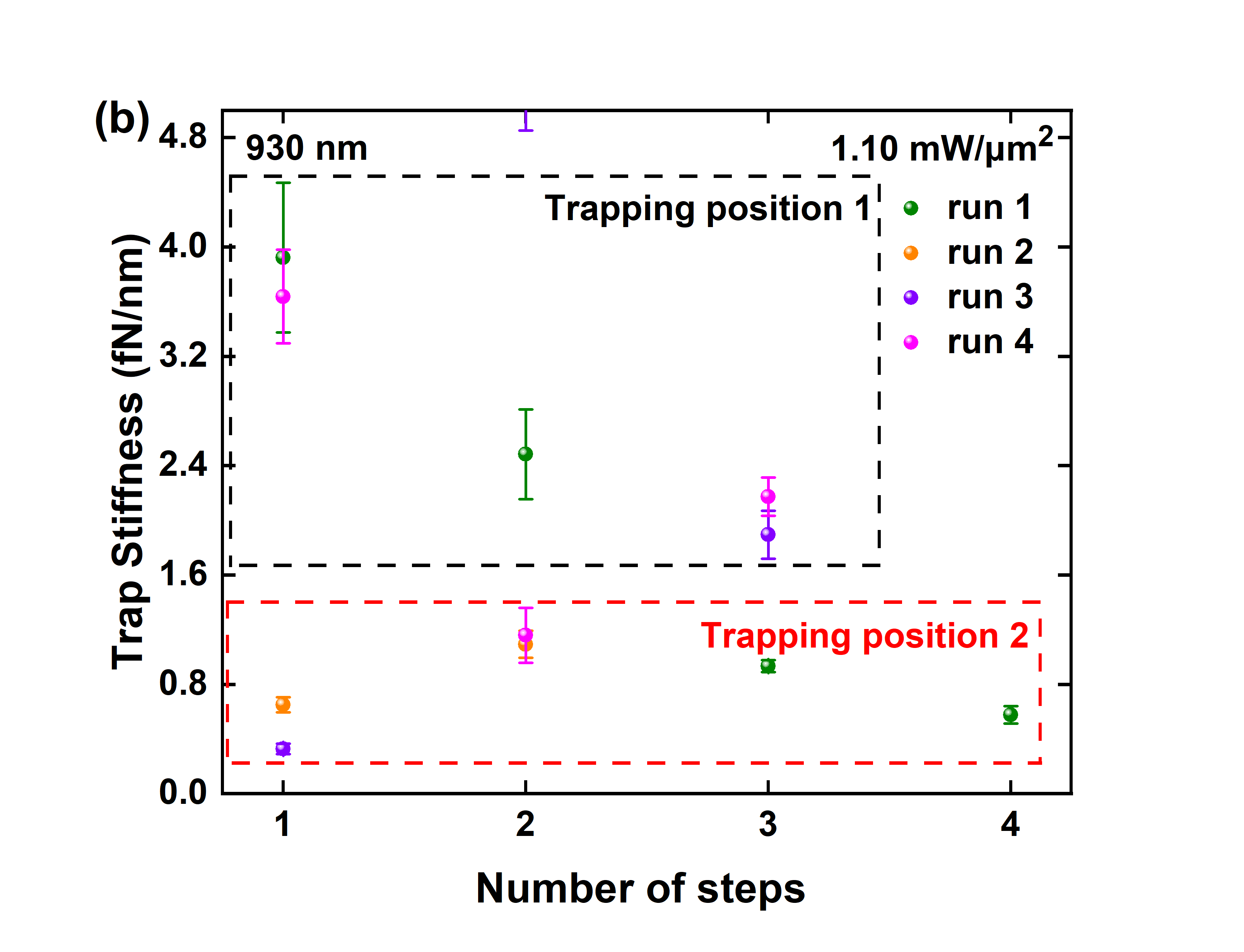}
\end{minipage}
\setlength\abovecaptionskip{0pt}
\caption{\label{Fig.4} Trap stiffness as a function of trapping events, \textit{i.e.}, number of steps or order of trapped nanoparticles, and multiple runs, for a trapping laser intensity of (a) 0.55~mW/$\mu$m$^{2}$ and (b) 1.10~mW/$\mu$m$^{2}$ near resonance at 930~nm. The dashed rectangles indicate  trapping positions of the ASR metamaterial where the particle may be trapped. The \textit{y}-error is the standard deviation of the trap stiffness measurements.}
\end{figure*}

\begin{equation}
    \tau = \frac{\gamma}{\kappa_{tot}}
\end{equation}
where~\textit{$\tau$} is the exponential decay time. We consider the Stoke drag coefficient and adjust it with the Faxen correction factor~\cite{R24,R22}. 

Figures~\ref{Fig.4} show the trap stiffness as a function of the trapping steps for four trapping runs (turning on and off the trapping beam) with near-resonance trap wavelength of  930~nm and a laser intensity of (a) 0.55~mW/$\mu$m$^{2}$ and (b) 1.10~mW/$\mu$m$^{2}$. To confirm single nanoparticle trapping, we analyzed data for comparable step heights (see Figure~\ref{Fig.3}(b)) for each observable trapping event. At a near-resonance wavelength and for low laser intensity, the first nanoparticle provides the maximum trap stiffness for the first trapping run and the third nanoparticle has maximum trap stiffness for the second run (Figure~\ref{Fig.4}(a)). In other words, the nanoparticle that is trapped at the hotspot where the optical force is maximum is not always the first particle to be trapped. Since the transportation of the nanoparticle due to Brownian motion is inherently very slow when the trapping laser is off, only  particles close to the excited plasmonic hotspot can be immobilized into plasmonic hotspots~\cite{R43}. We conclude that the nanoparticle which is closest to any one of the three plasmonic hotspots of the ASR unit (Figure~\ref{Fig.2}(a)) will be trapped at this specific hotspot under the influence of optical forces. We observe two distinct regions in which the trap stiffness magnitude is relatively constant (Figures~\ref{Fig.4}). It should be noted that a 20\% deviation of the actual particle's size was taken into account in the trap stiffness calculations. 
\begin{figure*}[htp]
\centering
\begin{minipage}[b]{0.33\linewidth}
\subfigure{\includegraphics[trim={1.7cm 0.5cm 1cm 1.8cm},clip, width=1\textwidth]{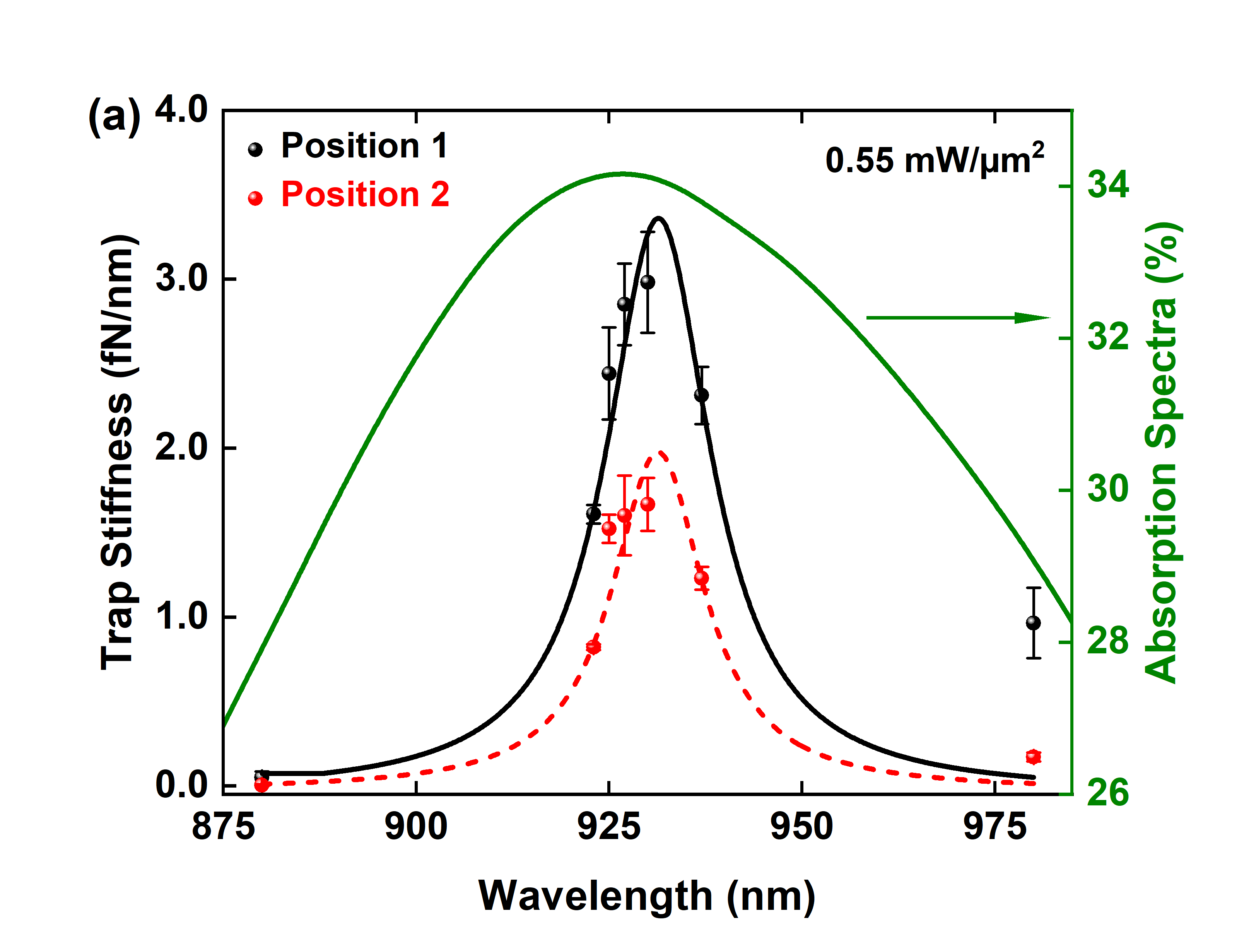}}
\end{minipage}\hfill
\begin{minipage}[b]{0.33\linewidth}
\subfigure{\includegraphics[trim={1.7cm 0.5cm 1cm 1.8cm},clip,width=1\textwidth]{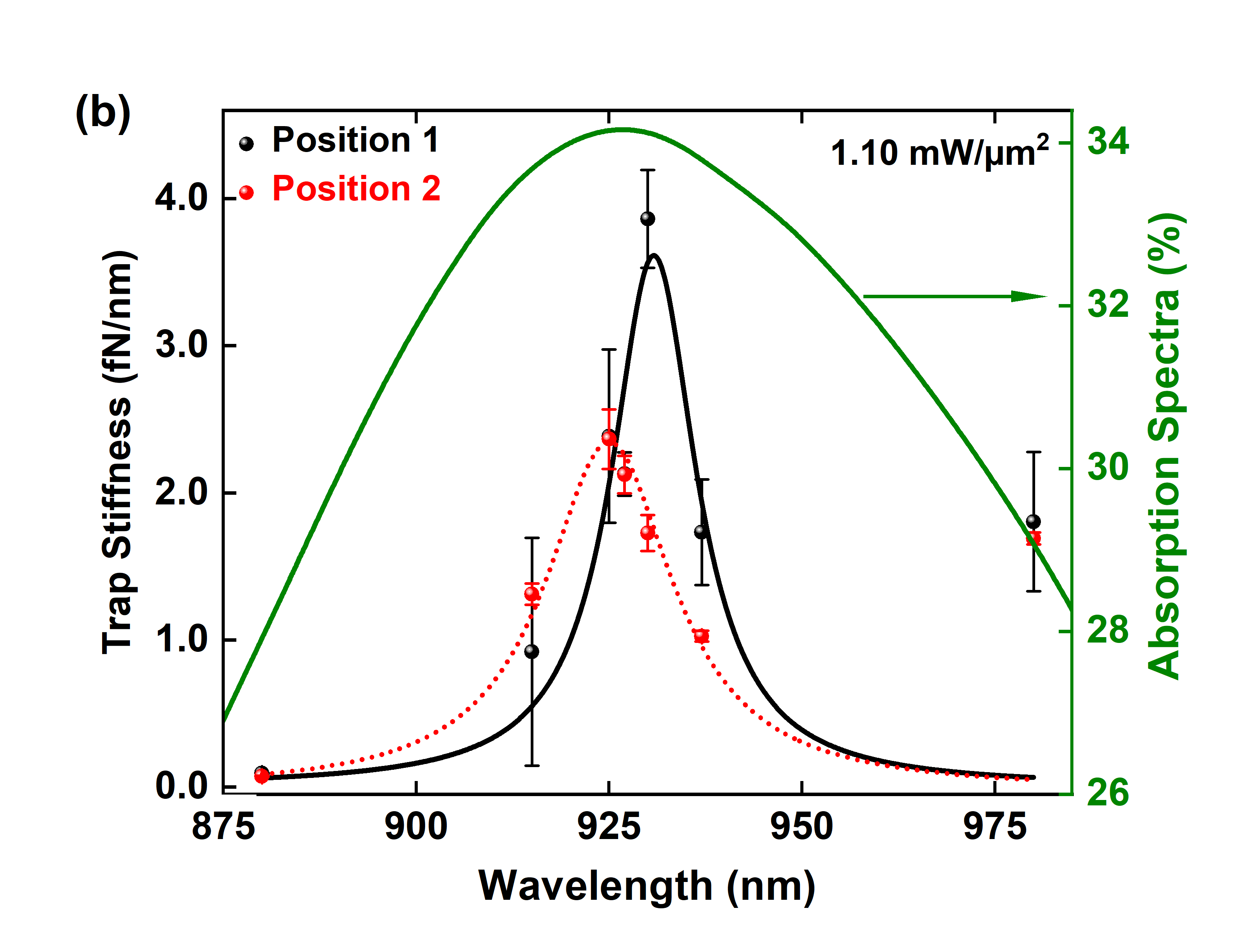}}
\end{minipage}\hfill
\begin{minipage}[b]{0.3\linewidth}
\subfigure{\includegraphics[trim={1.2cm 0.5cm 2.8cm 1.8cm},clip, width=1.01\textwidth]{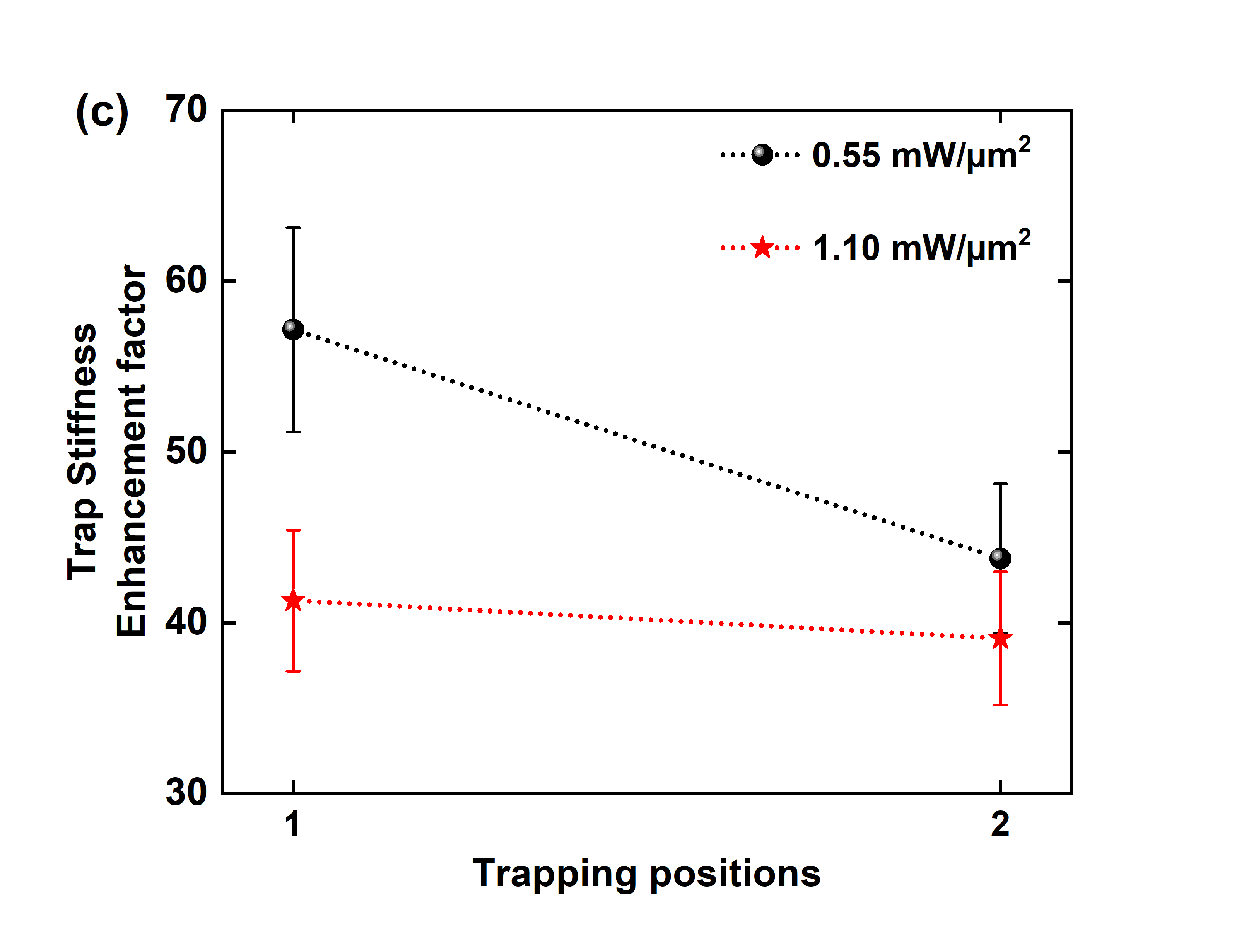}}
\end{minipage}
\setlength\abovecaptionskip{0pt}
\caption{\label{Fig.5} Absorption spectra (green line) and trap stiffness versus trapping wavelength and trapping positions for a laser intensity of (a) 0.55~mW/$\mu$$m^{2}$, and (b) 1.10~mW/$\mu$$m^{2}$. All red and black lines are Lorentz fitting result from the experimental data. The \textit{y}-error is the standard deviation of the trap stiffness measurements. (c) Trap stiffness enhancement factor as a function of trapping positions for low and high laser intensities.}
\end{figure*}

Figures~\ref{Fig.5}(a) and (b) show the trap stiffnesses for the two trapping positions (Position 1 and 2 (Figure~\ref{Fig.2} (a)) versus incident trapping wavelength for a low (0.55~mW/$\mu$m$^{2}$) and a high (1.10~mW/$\mu$m$^{2}$) incident laser intensity, respectively. The absorption spectrum is also plotted~\cite{R37}. The maximum trap stiffness is achieved for a trapping wavelength near resonance for both trapping positions. Unsurprisingly, the plasmonic hotspot with the weaker near-field (Position 2) provides a smaller trap stiffness compared with the strong near-field hotspot (Position 1). Using Lorentz fitting of the experimental trap stiffness values, we determined the trap stiffness peak for each trapping position at 931.4~nm for low incident trapping intensity and the result is in agreement with our previous observations~\cite{R37}. However, at a high (1.10~mW/$\mu$m$^{2}$) laser intensity, we observe a red-shift of the trap stiffness resonance peak as the particle moves from the weaker (Position 2, Figure~\ref{Fig.2}(a)) to the stronger (Position 1) hotspot as shown in  Figure~\ref{Fig.5}(b). This is in reasonable agreement with the theoretical optical force calculations, see Figure~\ref{Fig.2}(b). Moreover, we notice that the FWHM of the absorption spectra (green lines) are much broader than those for the trap stiffness curves for both low and high laser trapping intensities. This is in agreement with our previous observations~\cite{R37}.

Figure~\ref{Fig.5}(c) shows the trap stiffness enhancement factor versus trapping positions for both laser intensities. The trap stiffness enhancement is determined by normalizing the trap stiffness for each of the two trapping positions to that obtained at 880~nm, a wavelength far from the resonance peak.  The enhancement, which is due to the strong interaction between the trapped nanoparticle and the Fano resonance, decreases as the particle moves from trapping Position 1 to Position 2 for the higher laser intensity. Because the trap stiffness is directly proportional to the optical forces and to the near-field intensity, the trap stiffness enhancement factor follows the near-field strength of each plasmonic hotspot. This experimental feature indicates that our ASR metamaterial configuration can be used to identify the nanoparticle's location.

A common characteristic of all metallic nanostructures is absorption and ohmic losses in the metal, leading to Joule heating of the surrounding environment~\cite{R39,R48}. Photo-induced heat generation in metallic nanowires was investigated by using a thermal microscopy technique~\cite{R39}. The authors showed that the local distribution of the temperature in plasmonic nanostructures was fairly uniform despite the non-uniformity of the heat source density~\cite{R39}. This feature induces mismatch between the thermal hotspots which arise from the areas where charges can freely flow, and plasmonic hotspots from the tip effects and charge accumulation at the metal interface~\cite{R39}. Additionally, the local temperature increase results in thermally induced fluid convection, which exerts drag forces on suspended nanoparticles in a liquid environment~\cite{R49,R50}. These convection flows have improved the delivery of nanoparticles towards plasmonic hotspots where they can be trapped~\cite{R50,R49,R51}. Although the gradient forces exerted on a nanoparticle are strong at trapping Positions 1 (see Figure~\ref{Fig.2}(b)), due to their short-range the nanoparticle tends to be trapped at the hotspot closest to it. We can determine the specific position at which the nanoparticle will be trapped using the trap stiffness calculations. This sorting performance stems from the tunability of the potential wells and has applications in the trapping and sorting of biomolecules. However, to further investigate the origin of this mechanism, an analytical theoretical heating model of the metamaterial nanostructure would be required to explore, in depth, the contribution of thermally induced effects on the trapping performance as well as the particle's motion; it will be a focus of future work. 

The fact that the trap stiffness of each plasmonic hotspot (Figures~\ref{Fig.5}(a) and (b)) follows a similar trend as the corresponding optical gradient force (Figure~\ref{Fig.2}(b)) confirms the strong contribution of the Fano resonance to the trapping efficiency, as shown in our previous work~\cite{R37}. However, the collective heating of many hotspots, which  lead to an increase in the local temperature, results in convection effects. These photo-induced thermal effects may influence the trap stiffness enhancement factor at high laser intensities, as shown in Figure~\ref{Fig.5}(c). A certain range of laser intensities exists over which thermal effects are in synergy with the near-field, leading to  strengthening of the optical trap. The above-mentioned analysis indicates the extraordinary trapping ability of our configuration to demonstrate sequential trapping of 20 nm particles at low and high laser intensities. The exact nature of this trapping ability depends on several parameters and must be addressed, but it is beyond the scope of this work.
 
It remains a formidable challenge to sort out sub-wavelength particles with single-nanometer precision. Here, we have experimentally demonstrated a platform based on a chip consisting of an array of ASR metamaterial elements for multiple nanotrapping. We have studied the performance of our device by using trap stiffness measurements as a function of trapping position and trapping wavelength for both low and high laser intensities. We have shown that  subwavelength, dielectric nanoparticles can be transported to the plasmonic system by turning on and off the trapping laser beam when at near resonance. By using multiple plasmonic traps on a chip, controlled manipulation of nanoparticles and/or biomolecules for time periods lasting minutes in real-time is feasible.

\section{Data Availability}
The data that support the findings of this study are available from the authors upon reasonable request.

\begin{acknowledgments}
The authors would like to thank the Engineering Support Section at OIST for access to the nanofabrication facilities, Metin Ozer for technical assistance, and OIST Editing Section for reviewing the manuscript. This work was supported by funding from the Okinawa Institute of Science and Technology Graduate University. DGK acknowledges support from JSPS Grant-in-Aid for Scientific Research (C) Grant Number GD1675001.
\end{acknowledgments}

\bibliography{Bibliography}

\end{document}